\overfullrule=0pt
\input harvmac

\lref\BerkovitsBT{
  N.~Berkovits,
  ``Pure spinor formalism as an N=2 topological string,''
JHEP {\bf 0510}, 089 (2005).
[hep-th/0509120].
}

\lref\BerkovitsXU{
  N.~Berkovits,
  ``Quantum consistency of the superstring in AdS(5) x S**5 background,''
JHEP {\bf 0503}, 041 (2005).
[hep-th/0411170].
}

\lref\BerkovitsZK{
  N.~Berkovits,
  ``ICTP lectures on covariant quantization of the superstring,''
[hep-th/0209059].
}

\lref\BerkovitsFE{
  N.~Berkovits,
  ``Super Poincare covariant quantization of the superstring,''
JHEP {\bf 0004}, 018 (2000).
[hep-th/0001035].
}

\lref\MazzucatoJT{
  L.~Mazzucato,
  ``Superstrings in AdS,''
[arXiv:1104.2604 [hep-th]].
}

\lref\BH{
  N.~Berkovits and P.~S.~Howe,
  ``Ten-dimensional supergravity constraints from the pure spinor formalism for the superstring,''
Nucl.\ Phys.\ B {\bf 635}, 75 (2002).
[hep-th/0112160].
}

\lref\SiegelYD{
  W.~Siegel,
[arXiv:1005.2317 [hep-th]].
}

\def\bar{\overline}

\def\a{{\alpha}}

\def\ah{{\hat \a}}

\def\lh{{\widehat \lambda}}

\def\l{{\lambda}}
\def\lb{{\overline\lambda}}

\def\lb{{\overline\lambda}}
\def\b{{\beta}}

\def\g{{\gamma}}

\def\d{{\delta}}
\def\e{{\epsilon}}

\def\L{{\Lambda}}
\def\Lb{{\bar\Lambda}}

\def\half{{1\over 2}}
\def\p{{\partial}}

\def\t{{\theta}}

\Title{\vbox{\baselineskip12pt
\hbox{}}}
{{\vbox{\centerline{Twistor Origin of the Superstring
}}} }
\bigskip\centerline{Nathan Berkovits\foot{e-mail: nberkovi@ift.unesp.br}}
\bigskip
\centerline{\it ICTP South American Institute for Fundamental Research 
}
\centerline{\it Instituto de F\'\i sica Te\'orica, UNESP - Univ. 
Estadual Paulista }
\centerline{\it Rua Dr. Bento T. Ferraz 271, 01140-070, S\~ao Paulo, SP, Brasil}
\bigskip

\vskip .3in

After introducing a d=10 pure spinor $\lambda^\alpha$, the Virasoro constraint $\partial x^m \partial x_m =0$ can be replaced
by the twistor-like constraint $\partial x^m (\gamma_m \lambda)_\alpha=0$. Quantizing
this twistor-like constraint leads to the pure spinor formalism for the superstring where the
fermionic superspace variables $\theta^\alpha$ and their conjugate momenta come from
the ghosts and antighosts of the twistor-like constraint.

\vskip .3in

\Date {September 2014}

\newsec{Introduction}

The conventional manner to obtain the superstring from the bosonic string is to
generalize the worldsheet conformal invariance generated by the Virasoro constraint
$\partial x^m \partial x_m =0$ to the worldsheet N=1 superconformal invariance generated
by the super-Virasoro constraint $\partial x^m \psi_m=0$ where $\psi^m$ is the
fermionic worldsheet superpartner of $x^m$. This Ramond-Neveu-Schwarz (RNS) construction of the superstring \ref\RNS{
  P.~Ramond,
  ``Dual Theory for Free Fermions,''
Phys.\ Rev.\ D {\bf 3}, 2415 (1971)\semi
A.~Neveu and J.~H.~Schwarz,
 ``Factorizable dual model of pions,''
Nucl.\ Phys.\ B {\bf 31}, 86 (1971).} was developed in the 1970's, and although it is based on 
the simple geometrical idea of generalizing conformal invariance to superconformal
invariance, it has the disadvantage that spacetime supersymmetry is only present
after including both periodic and antiperiodic conditions for $\psi^m$ and performing a GSO projection \ref\gso{F.~Gliozzi, J.~Scherk and D.~I.~Olive,
  ``Supersymmetry, Supergravity Theories and the Dual Spinor Model,''
Nucl.\ Phys.\ B {\bf 122}, 253 (1977).} which truncates out states constructed from an even number of $\psi^m$ variables. This lack of manifest spacetime supersymmetry makes it difficult to compute
scattering amplitudes involving fermionic states and has prevented the RNS
formalism from being used to describe Ramond-Ramond backgrounds.

In the 1980's, Green and Schwarz developed a new formalism \ref\greens{M.~B.~Green and J.~H.~Schwarz,
  ``Supersymmetrical Dual String Theory,''
Nucl.\ Phys.\ B {\bf 181}, 502 (1981)\semi M.~B.~Green and J.~H.~Schwarz,
  ``Covariant Description of Superstrings,''
Phys.\ Lett.\ B {\bf 136}, 367 (1984).} for the superstring
in which spacetime supersymmetry is manifest and is constructed using a spacetime spinor variable $\theta^\alpha$ intead of the spacetime vector variable $\psi^m$ of the
RNS formalism. In addition to worldsheet conformal invariance, their superstring
action contains a fermionic symmetry called ``kappa symmetry'' \ref\siegk{W.~Siegel,
 ``Hidden Local Supersymmetry in the Supersymmetric Particle Action,''
Phys.\ Lett.\ B {\bf 128}, 397 (1983).} which replaces the
N=1 worldsheet superconformal invariance of the RNS formalism. However, the structure of kappa symmetry prevented quantization of the Green-Schwarz formalism except in light-cone gauge, which complicates the
computation of scattering amplitudes and the quantum description of Ramond-Ramond
backgrounds.

Starting in 2000, a new formalism for the superstring has been developed in which
spacetime supersymmetry is manifest and which can be easily quantized in a covariant 
manner \BerkovitsFE. In addition to the fermionic spinor variable $\theta^\alpha$ of the
Green-Schwarz formalism, this new formalism includes a bosonic spinor variable $\lambda^\alpha$ which satisfies the d=10 ``pure spinor'' constraint 
\eqn\pureintro{\l \gamma^m\l =0}
for
$m=0$ to 9. Unlike the RNS and Green-Schwarz formalisms, it has been successfully used to compute multiloop amplitudes involving both bosonic and fermionic states \ref\mafra{ H.~Gomez and C.~R.~Mafra,
  ``The closed-string 3-loop amplitude and S-duality,''
JHEP {\bf 1310}, 217 (2013).
[arXiv:1308.6567 [hep-th]]\semi  C.~R.~Mafra, O.~Schlotterer and S.~Stieberger,
  ``Complete N-Point Superstring Disk Amplitude I. Pure Spinor Computation,''
Nucl.\ Phys.\ B {\bf 873}, 419 (2013).
[arXiv:1106.2645 [hep-th]]\semi H.~Gomez and C.~R.~Mafra,
  ``The Overall Coefficient of the Two-loop Superstring Amplitude Using Pure Spinors,''
JHEP {\bf 1005}, 017 (2010).
[arXiv:1003.0678 [hep-th]].}
and to prove the quantum consistency of Ramond-Ramond backgrounds such
as $AdS_5\times S^5$ \ref\quantumads{N.~Berkovits,
  ``Quantum consistency of the superstring in AdS(5) x S**5 background,''
JHEP {\bf 0503}, 041 (2005).
[hep-th/0411170].\semi A.~Mikhailov and S.~Schafer-Nameki,
  ``Perturbative study of the transfer matrix on the string worldsheet in AdS(5) x S**5,''
Adv.\ Theor.\ Math.\ Phys.\  {\bf 15}, 913 (2011).
[arXiv:0706.1525 [hep-th]]\semi
 L.~Mazzucato,
  ``Superstrings in AdS,''
Phys.\ Rept.\  {\bf 521}, 1 (2012).
[arXiv:1104.2604 [hep-th]].}.

However, unlike the RNS formalism, the geometric origin of this new formalism was not
understood. Physical states and scattering amplitudes are defined using a gauge-fixed action and a nilpotent BRST operator $Q$ constructed 
from the Green-Schwarz variables and the pure spinor $\lambda^\alpha$ as
\eqn\qintro{Q = \int dz~\lambda^\a d_\a,}
where $d_\a$ is the fermionic Green-Schwarz-Siegel constraint \ref\siegelclassical{W.~Siegel,
  ``Classical Superstring Mechanics,''
Nucl.\ Phys.\ B {\bf 263}, 93 (1986).} which generates kappa
symmetry. But despite several attempts \ref\brstattempts{M.~Matone, L.~Mazzucato, I.~Oda, D.~Sorokin and M.~Tonin,
  ``The Superembedding origin of the Berkovits pure spinor covariant quantization of superstrings,''
Nucl.\ Phys.\ B {\bf 639}, 182 (2002).
[hep-th/0206104]\semi Y.~Aisaka and Y.~Kazama,
  ``Origin of pure spinor superstring,''
JHEP {\bf 0505}, 046 (2005).
[hep-th/0502208]\semi N.~Berkovits,
  ``Explaining the Pure Spinor Formalism for the Superstring,''
JHEP {\bf 0801}, 065 (2008).
[arXiv:0712.0324 [hep-th]]\semi N.~Berkovits,
  ``Pure spinors, twistors, and emergent supersymmetry,''
JHEP {\bf 1212}, 006 (2012).
[arXiv:1105.1147 [hep-th]].}, this pure spinor BRST operator was not obtained in a
simple manner by gauge-fixing a worldsheet reparameterization invariant action.

In this paper, an elegant geometrical origin for this formalism will be proposed
and the pure spinor BRST operator of \qintro\ will be obtained by gauge-fixing
a simple worldsheet reparameterization invariant action. Surprisingly, this
reparameterization invariant action will be constructed entirely from bosonic worldsheet
variables, and the fermionic worldsheet variables $\t^\a$ and their conjugate momenta
will come from ghosts and antighosts associated with the gauge fixing.

The bosonic variables in the worldsheet action will consist of the usual d=10 spacetime vector
variable $x^m$ together with a spacetime spinor variable $\l^\a$ satisfying the d=10 pure spinor
constraint $\l\g^m\l=0$. The pure spinor constraint implies that only 11 of the 16 components of $\l^\a$ are independent, and after Wick-rotation
to Euclidean signature, $\l^\a$ parameterizes the eleven-dimensional complex space ${{SO(10)}\over{U(5)}}\times C^*$ where $C^*$ is the complex plane minus the origin 
\ref\nekrasov{ 
  N.~A.~Nekrasov,
  ``Lectures on curved beta-gamma systems, pure spinors, and anomalies,''
[hep-th/0511008].}.

Instead of generalizing the Virasoro constraint $T=-\half \p x^m \p x_m=0$ to a super-Virasoro
constraint as in the RNS formalism, the Virasoro constraint $T=0$ will instead by replaced
by the twistor-like constraint
\eqn\twistc{C_\a = -\half \p x^m (\g_m\l)_\a =0.}
Note that $C_\a =0$ implies $T=0$ since $T$ is equal to ${1\over{\l^\b\lb_\b}}C_\a (\g_m\lb)^\a \p x^m$
where $\lb_\a$ is any spinor satisfying $\l^\a \lb_\a \neq 0$. As discussed by several authors
\ref\pstw{    L.P. Hughston, ``The Wave Equation in Even
Dimensions,'' in Further Advances in Twistor Theory, vol. 1,
Research Notes in Mathematics 231, Longman, 26-27 (1990)\semi
L.P. Hughston, ``A Remarkable Connection between the Wave Equation
and Pure Spinors in Higher Dimensions,'' in Further Advances in
Twistor Theory, vol. 1, Research Notes in Mathematics 231,
Longman, 37-39 (1990) \semi L.~P.~Hughston and L.~J.~Mason,
  ``A Generalized Kerr-Robinson Theorem,''
Class.\ Quant.\ Grav.\  {\bf 5}, 275 (1988)\semi J.~P.~Harnad and S.~Shnider,
  ``Isotropic Geometry, Twistors And Supertwistors. 1. The Generalized Klein Correspondence And Spinor Flags,''
J.\ Math.\ Phys.\  {\bf 33}, 3197 (1992)\semi J.~P.~Harnad and S.~Shnider,
 ``Isotropic geometry and twistors in higher dimensions. 2: Odd dimensions, reality conditions, and twistor superspaces,''
J.\ Math.\ Phys.\  {\bf 36}, 1945 (1995)\semi P. Budinich and
A. Trautman, The Spinorial Chessboard (Trieste Notes in Physics,
Springer-Verlag, Berlin, 1988)
\semi
P. Budinich, ``From the geometry of pure spinors with their
division algebras to fermion's physics,'' Found. Phys. 32, 1347 (2002),
hep-th/0107158\semi P. Furlan and R.
Raczka, ``Intrinsic nonlinear spinor wave equations
associated with nonlinear spinor representations,''
Journal of Mathematical Physics 27, 1883 (1986)\semi  P.Furlan and R.Raczka,
``Nonlinear  Spinor Representations,''
Journal of Mathematical Physics 26, 3021 (1985)\semi N.~Berkovits and S.~A.~Cherkis,
  ``Higher-dimensional twistor transforms using pure spinors,''
JHEP {\bf 0412}, 049 (2004).
[hep-th/0409243].}, pure
spinors are the natural generalization to higher dimensions of d=4 Penrose twistors \ref\penrose{R.~Penrose,
  ``Twistor algebra,''
J.\ Math.\ Phys.\  {\bf 8}, 345 (1967).},
and \twistc\ is the d=10 stringy version \ref\hughstonshaw{E. Witten, ``Twistor-like transform in ten dimensions,'' Nucl. Phys. B266, 245
(1986)\semi 
B.~E.~W.~Nilsson,
  ``Pure Spinors as Auxiliary Fields in the Ten-dimensional Supersymmetric {Yang-Mills} Theory,''
Class.\ Quant.\ Grav.\  {\bf 3}, L41 (1986)\semi
L.~P.~Hughston and W.~T.~Shaw,
  ``Real Classical Strings,''
Proc.\ Roy.\ Soc.\ Lond.\ A {\bf 414}, 415 (1987)\semi L.~P.~Hughston and W.~T.~Shaw,
  ``Classical Strings In Ten-dimensions,''
Proc.\ Roy.\ Soc.\ Lond.\ A {\bf 414}, 423 (1987)\semi A. Bengtsson, I. Bengtsson, M. Cederwall and N. Linden,
``Particles, superparticles and twistors,'' Phys. Rev. D36, 1766 (1987)\semi
E. Sokatchev, ``Harmonic superparticle,'' Class. Quant. Grav. 4, 237 (1987)\semi
D. Sorokin, V. Tkach and D. Volkov, ``Superparticles,
twistors and Siegel symmetry,'' Mod. Phys. Lett. A4, 901 (1989)\semi
N. Berkovits, ``A supertwistor description of the massless superparticle
in ten-dimensional superspace,'' Nucl. Phys. B350, 193 (1991)\semi
E. Bergshoeff, P. Howe, C. Pope, E. Sezgin
and E. Sokatchev, ``Ten-dimensional supergravity from lightlike
integrability in loop superspace,'' Nucl. Phys. B354, 113 (1991)\semi 
P. Howe, ``Pure spinors, function superspaces and supergravity
theories in ten-dimensions and eleven-dimensions,'' Phys. Lett. B273, 90
(1991)\semi P. Howe, ``Pure spinors lines in superspace and
ten-dimensional
supersymmetric theories,'' Phys. Lett. B258, 141 (1991).
 } of the d=4 twistor constraint $({{\p}\over{\p\tau}}x_{a\dot a}) \l^a =0$
where $a,\dot a =1$ to 2 and $x_{a\dot a}(\tau)$ is a d=4 light-like trajectory.

The worldsheet reparameterization invariant action for $x^m$ and $\l^\a$ will be
\eqn\wsa{S = \int d^2 z (\det{e})(\half\nabla x^m \bar\nabla x_m + w_\a \bar\nabla \l^\a +
L^\a C_\a + \l^\a \Lb_\a)}
where $\nabla = e_-^J\p_J$, $\bar\nabla = e_+^J \p_J$, $e_\pm^J$ is the usual two-dimensional vierbein, and $L^\a$ is a Lagrange multiplier which enforces the
constraint $C_\a =0$. In addition, the term
$\l^\a \Lb_\a$ has been included in the Lagrangian where $\bar \L_\a$ is a bosonic
pure spinor of opposite chirality to $\l^\a$. If $\Lb_\a$ is interpreted (in Euclidean signature)
as the complex conjugate to $\l^\a$, the term $\l^\a \Lb_\a$ concentrates the functional
integration over $\l^\a$ to the region near $\l^\a =0$ and eliminates the divergence coming from
functional integration over the non-compact zero modes of $\l^\a$.

To quantize this action, one needs to gauge-fix the invariances generated by the
constraint $C_\a$ of \twistc. But because $\l\g^m\l=0$, only 5 of the 16 components of $C_\a$
are independent. To gauge-fix, one should first restrict $\l^\a$ to a patch of pure spinor
space where $\lb_\a \l^\a \neq 0$ for some fixed constant pure spinor $\lb_\a$. On this
patch of pure spinor space, one can restrict the Lagrange multiplier $L^\a$ to satisfy the 11 independent constraints
$L\g^{mn}\lb=0$, and the remaining 5 components of $L^\a$ can be gauge-fixed in
the usual manner to produce 5 fermionic Faddeev-Popov ghosts and antighosts, $f^\a$
and $m_\a$, which satisfy the constraints $f\g^{mn}\lb =0$ and $m\g^{mn}\l=0$.

On this same patch of pure spinor space, one can similarly gauge-fix $\Lb_\a$ to be
proportional to the constant pure spinor $\lb_\a$. This gauge-fixing procedure produces
additional fermionic Faddeev-Popov ghosts and antighosts, $g_\a$ and $n^\a$, which
because of the pure spinor constraint on $\Lb_\a$, are constrained to satisfy
$g\g^m\lb =0$ and $n\g^m\l=0$ so they each have 11 independent components.

Note that there are no additional Faddeev-Popov ghosts and antighosts coming from gauge-fixing worldsheet reparameterization
invariance to conformal gauge since the Virasoro constaint $T=0$ is already implied
by the twistor-like constraint of \twistc. This explains why the $b$ ghost satisfying $\{Q,b\}=T$
is not a fundamental worldsheet variable in the pure spinor formalism, but is a composite
operator constructed out of the other variables.

Although the constraints on the fermionic ghosts $(f^\a, g_\a)$ and antighosts $(m_\a, n^\a)$
depend on the choice of patch of pure spinor space, one can define an unconstrained fermionic
spinor variable $\t^\a$ and its conjugate momentum $p_\a$ as
\eqn\tptp{\t^\a = f^\a + n^\a \quad {\rm and} \quad p_\a = g_\a + m_\a}
which are independent of the choice of $\lb_\a$. In terms of these unconstrained fermionic
variables, the Faddeev-Popov ghost contribution to the action is 
\eqn\fpa{\int d^2 z (m_\a \bar\p f^\a + n^\a \bar\p g_\a) =\int d^2 z~ p_\a\bar\p\t^\a,}
so the gauge-fixed action is 
\eqn\ggff{S = \int d^2 z (\half\p x^m \bar\p x_m + w_\a \bar\p\l^\a + p_\a \bar\p\t^\a).}
And the resulting BRST operator is
\eqn\brr{Q = \int dz (\l^\a p_\a + C_\a \t^\a -{1\over 8} (\l\g^m\t)(\t\g_m\p\t)) = \int dz ~\l^\a d_\a,}
where $d_\a$ is the supersymmetric Green-Schwarz-Siegel constraint \siegelclassical\ and the term
$-{1\over 8}(\l\g^m\t)(\t\g_m\p\t)$ in $Q$ comes from the non-abelian constraint algebra
$[C_\a, C_\b] = {1\over 8} (\g^m\l)_{[\a} (\g_m\nabla\l)_{\b]} $. 

So the gauge-fixed action and BRST operator of the pure spinor formalism are obtained
by quantizing the simple worldsheet reparameterization invariant action of \wsa. Since the fermionic worldsheet variables in the gauge-fixed action come from Faddeev-Popov ghosts and antighosts, a natural question is how fermionic variables can appear in light-cone gauge where ghosts are absent. This question can be studied
by simplifying to the d=10 superparticle \ref\bsuperp{ N.~Berkovits,
  ``Covariant quantization of the superparticle using pure spinors,''
JHEP {\bf 0109}, 016 (2001).
[hep-th/0105050].} where the twistor-like constraint of \twistc\ reduces
to $C_\a =-\half P^m (\g_m\l)_\a=0$. Although $C_\a$ has 5 independent components, it implies a single mass-shell constraint $P^2=0$ for the $x^m$ dependence. And since $\l^\a$ dependence is fixed by the $\l^\a\Lb_\a$ term in the Lagrangian to be near $\l^\a=0$, there are 4 components of $C_\a$ which overconstrain the classical worldsheet variables. These 4 extra constraints of $C_\a$ lead
to 4 fermionic variables together with their conjugate momenta which are the usual 8 light-cone Green-Schwarz fermions.

It is interesting to point out that this same phenomenon occurs for the d=11 pure spinor
description of the superparticle \ref\bmem{M.~Cederwall, B.~E.~W.~Nilsson and D.~Tsimpis,
  ``Spinorial cohomology and maximally supersymmetric theories,''
JHEP {\bf 0202}, 009 (2002).
[hep-th/0110069]\semi N.~Berkovits,
  ``Towards covariant quantization of the supermembrane,''
JHEP {\bf 0209}, 051 (2002).
[hep-th/0201151]\semi  M.~Cederwall,
  ``D=11 supergravity with manifest supersymmetry,''
Mod.\ Phys.\ Lett.\ A {\bf 25}, 3201 (2010).
[arXiv:1001.0112 [hep-th]]\semi M.~Cederwall and A.~Karlsson,
  ``Loop amplitudes in maximal supergravity with manifest supersymmetry,''
JHEP {\bf 1303}, 114 (2013).
[arXiv:1212.5175 [hep-th]].}
which describes d=11 supergravity. In this case, the
bosonic variables are $x^M$ for $M=0$ to 10 and $\l^A$ for $A=1$ to 32 where $\l^A$ satisfies
the pure spinor constraint $\l\g^M\l=0$ that reduces its 32 components to 23 independent components. The twistor-like constraint $C_A =-\half P^M (\g_M\l)_A=0$ has 9 independent
components, and implies the d=11 mass-shell constraint $P^2=0$. So there are 8 components
of $C_A$ which overconstrain the classical variables and lead to 8 fermionic variables
and their conjugate momenta in light-cone gauge.

After describing the worldsheet reparameterization invariant 
action of \wsa\ and its gauge invariances in sections 2 and 3 of this paper, the gauge-fixing
procedure on a patch where $\lb_\a\l^\a\neq 0$ will be discussed in section 4. In section 5,
the ``minimal'' version of the pure spinor formalism will be derived using this procedure, and in section 6, the ``non-minimal'' version of the pure spinor formalism \ref\nonmin{N.~Berkovits,
  ``Pure spinor formalism as an N=2 topological string,''
JHEP {\bf 0510}, 089 (2005).
[hep-th/0509120]\semi N.~Berkovits and N.~Nekrasov,
  ``Multiloop superstring amplitudes from non-minimal pure spinor formalism,''
JHEP {\bf 0612}, 029 (2006).
[hep-th/0609012].} will be derived
by upgrading $\lb_\a$ from a constant pure spinor to a worldsheet variable.

Finally, 
a conjecture for generalizing this procedure to curved Type II supergravity backgrounds including Ramond-Ramond fields will be proposed in section 7. Since all fermionic variables in
the worldsheet action arise from Faddeev-Popov ghosts, Ramond-Ramond background
fields will not directly appear in the reparameterization invariant action and will only appear after performing the gauge-fixing procedure. The absence
of physical Ramond-Ramond fields from
the classical action implies that there is non-trivial
BRST cohomology at nonzero ghost number where the ghosts 
$(f^\a, g_\a)$ and antighosts $(m_\a, n^\a)$ are defined to carry ghost-number
$+1$ and $-1$. This fact is not surprising since the patch-independent variables
$\t^\a$ and $p_\a$ of \tptp\ do not have well-defined ghost number when 
ghost number is defined
in terms of $(f^\a, g_\a)$ and $(m_\a, n^\a)$.

\newsec{Worldsheet action}

The worldsheet variables in the reparameterization invariant action will include the
spacetime $x^m$ variables ($m=0$ to 9), the left-moving bosonic pure spinor $\l^\a$
variables ($\a=1$ to 16) and their conjugate momenta $w_\a$, and the right-moving
bosonic pure spinor $\widehat\l^\ah$ variables and their conjugate momenta $\widehat w_\ah$.
Because of the pure spinor constraints 
\eqn\pures{\l\g^m\l = \lh\g^m\lh =0,}
the conjugate
momenta $w_\a$ and $\widehat w_\ah$ can only appear in combinations which are
invariant under the gauge transformations $\d w_\a = c^m (\g_m\l)_\a$ and $ \d \widehat w_\ah  = \widehat c^m (\g_m\lh)_\ah$
for arbitrary $c^m$ and $\widehat c^m$.
Note that $(x^m, \l^\a, \lh^\ah)$ are worldsheet scalars, and $w_\a$ and $\widehat w_\ah$
carry conformal weight $(1,0)$ and $(0,1)$ respectively.
For the Type IIA (or Type IIB) superstring, the $\hat \a$ index on right-moving spinors
denotes the opposite (or same) spacetime chirality as the unhatted $\a$ index on left-moving
spinors. And the heterotic superstring is obtained by replacing the right-moving sector with
the same right-moving sector as in the RNS heterotic formalism.

The Type II worldsheet action in a flat background is
\eqn\action{
S = \int d^2 z ~(\det e) [\half\nabla x^m \bar\nabla x_m + w_\a \bar\nabla \lambda^\a
+ \widehat w_\ah \nabla \lh^\ah}
$$+ L^\a C_\a + \bar\Lambda_\a \l^\a + \widehat L^\ah \widehat C_\ah + \widehat {\bar\Lambda}_\ah \lh^\ah +{1\over 4} (L\g^m\l)(\widehat L\g_m\lh) ] $$
where $\nabla = e_-^J \p_J $, $\bar\nabla = e_+^J\p_J$, $e_\pm^J$ is the worldsheet vielbein for $J=1$ to 2, 
$C_\a$ and $\widehat C_\ah$ are the twistor-like constraints
\eqn\twistorline{ C_\a= -\half \nabla x^m (\g_m \l)_\a, \quad \widehat C_\ah = -\half\bar\nabla x^m (\g_m\lh)_\ah,}
$L^\a$ and $\widehat L^\ah$ are Lagrange multipliers of conformal weight $(0,1)$ and $(1,0)$,
and $\bar\Lambda_\a$ and $\widehat {\bar\Lambda}_\ah$ are Lagrange multipliers of conformal weight $(1,1)$.

Just as $\l^\a$ and $\lh^\ah$ are pure spinors satisfying the constraint of \pures,
the Lagrange multipliers $\bar\L_\a$ and $\widehat{\bar\L}_\ah$ will also be required to be
pure spinors satisfying the constraints
\eqn\purel{\bar\L\g^m\bar\L = {\widehat{\bar\L}}\g^m\widehat{\bar\L} =0,}
so that $\bar\L_\a$ and $\widehat{\bar\L}_\ah$ each have 11 independent complex components.
After Wick rotation to Euclidean signature, pure spinors parameterize the complex space 
${{SO(10)}\over{U(5)}}\times C^*$ where $C^*$ denotes the complex plane minus the origin.
So all components of a pure spinor cannot be simultaneously zero.
To globally paramaterize pure spinors, one therefore needs to divide the space into 16 patches ${\cal O}_\a$ for $\a=1$ to 16 where, on the patch ${\cal O}_\a$, the component $\l^\a$ and $\bar\L_\a$ of the pure spinors are required to be nonvanishing \nekrasov. 

In addition to acting as a Lagrange multiplier for the nonzero modes of $\l^\a$, the zero modes of $\Lb_\a$ can be interpreted as a regulator for the zero modes of $\l^\a$. In other words,
if the zero modes of $\Lb_\a$ are interpreted (after Wick rotation) as the complex conjugate
of the $\l^\a$ zero modes, the term $\l^\a\Lb_\a$ in the action acts as a Gaussian regulator
for the functional integration over these non-compact pure spinor zero modes.
Note that the pure spinor constraint on $\bar\L_\a$ implies that it cannot be used
to remove all $\l^\a$ dependence from the action of \action. For example, the shift
\eqn\shift{\bar\L_\a \to \bar\L_\a + \bar\nabla w_\a +\half \nabla x^m (\g_m L)_\a}
which would naively remove $\l^\a$ dependence from the action is not allowed since it does not preserve
the constraint of \purel. 

To simplify notation, the right-moving sector will be ignored for the rest of this paper when it plays an identical role to the left-moving sector. 

\newsec{Gauge Invariances}

Because of the first-class constraint $C_\a =-\half \nabla x^m (\g_m\l)_\a$, the worldsheet
action of \action\ is invariant under the gauge transformation
\eqn\twistorg{\d x^m =\half \l\g^m f,\quad \d w_\a =-\half \nabla x^m (\g_m f)_\a +
{1\over 4} (\widehat L\g^m \lh)(\g_m f)_\a, \quad 
\d L^\a = \bar\nabla f^\a,}
\eqn\deltag{\delta \bar\L_\a ={1\over{16 (\l\Lb)}} (\g^m\g^n\bar\L)_\a 
[\nabla(\l\g_m f) (\l\g_n L) - \nabla (\l\g_m L)(\l\g_n f)],}
where $f^\a$ is an arbitrary infinitesimal parameter and the variation of $\d\Lb_\a$ is necessary since $[C_\a, C_\b] ={1\over 8} (\g_m\l)_{[\a} (\g_n\nabla\l)_{\b]}$ implies that
\eqn\implc{\l^\a \d \bar\Lambda_\a ={1\over 8} [(\l\g^m L)\nabla(\l \g_m f) - (\l\g^m f)\nabla(\l \g_m L)].} 
Although \implc\ does not uniquely
determine \deltag, it will be later argued that any other $\d\Lb_\a$ that satisfies \implc\
will lead to the same BRST operator up to a similarity transformation.

The gauge invariance $x^m \sim x^m +\half \l\g^m f $ of \twistorg\ is the d=10 generalization
of the d=4 twistor symmetry \penrose
\eqn\fourt{x^{a\dot a} \sim x^{a \dot a} + \l^a f^{\dot a} \quad{\rm where ~~~} a,\dot a = 1~~{\rm to}~~2}
that identifies points on a self-dual plane and leaves the twistor variable 
$\mu^{\dot a} = x^{a \dot a} \l_a$ invariant. So as discussed in \pstw, the d=10 pure spinor
variable $\l^\a$ plays a similar role to the d=4 twistor variable $\l^a$ of Penrose.

The worldsheet action of \action\ is also invariant under the gauge transformation generated
by $\l^\a$ which is
\eqn\gaugel{\d w_\a = g_\a, \quad \d \Lb_\a =\bar\nabla g_\a + {1\over{2(\l\Lb)}} (g \g^m \bar\nabla\Lb) (\g_m \l)_\a,}
where $g_\a$ is an arbitrary infinitesimal parameter of conformal weight $(1,0)$ satisfying
$(g\g^{m}\Lb)=0$ and the second term in $\d\Lb_\a$ is needed so that $\d\Lb\g^m\Lb =0$.
Furthermore, since $\l\g^m\l=0$, \action\ is invariant under the gauge transformations
\eqn\gaugeL{\d L^\a = c_{mn} (\g^{mn}\l)^\a}
for arbitrary $c^{mn}$, which implies that 11 of the 16 components of $L^\a$ can be
gauged away.

Finally, the worldsheet action is invariant under the usual worldsheet reparameterizations
generated by the Virasoro constraint 
\eqn\stress{T = -\half\nabla x^m \nabla x_m - w_\a \nabla \l^\a.}
However, these reparameterizations are already included as a special case of the previous
gauge transformations. This can be seen from the fact that the Virasoro constraint of \stress\
can be expressed as a linear combination of the other constraints $C_\a=-\half\nabla x^m (\g_m\l)_\a$ and $\l^\a$ as
\eqn\lc{ T = C_\a {{\nabla x^m (\g_m \Lb)^\a}\over{(\l\Lb)}} + 
\nabla \l^\a {{(\l\g_{mn}w)(\g^{mn}\Lb)_\a +2 (\l w)\Lb_\a}\over{8(\l\Lb)}}.}
So all dependence of the action of \action\ on off-diagonal components of the worldsheet
vierbein can be removed by an appropriate shift of the Lagrange multipliers $(L^\a,\Lb_\a)$
and $(\widehat L^\ah, \widehat{\bar\Lambda}_\ah)$.

\newsec{Gauge Fixing}

After shifting the Lagrange multipliers to eliminate the off-diagonal components of the worldsheet
vierbien, the worldsheet action can be expressed in
conformal gauge where $e_{\pm}^J$ is proportional to $\d_\pm^J$ so that 
$\nabla \to \p$ and $\bar\nabla \to \bar\p$. One then needs to fix the gauge invariances
of  \twistorg,  \gaugel\ and \gaugeL. The first step to perform this gauge fixing is to 
restrict the pure spinor 
$\l^\a$ to a patch ${\cal O}_\a$ where one of its components is required to
be nonzero. This patch can be defined by introducing a {\it constant} pure spinor
$\bar\l_\a$ and requiring that $\bar\l_\a \l^\a$ is nonzero on the patch. Different choices
of the constant pure spinor $\bar\l_\a$ correspond to different patches ${\cal O}_\a$,
and consistency of the gauge fixing will require that the resulting gauge-fixed action and
BRST operator are independent of the choice of $\bar\l_\a$.

On the patch where $\bar\l_\a \l^\a$ is nonzero, the gauge invariance of \gaugeL\ implies
that one can gauge fix $L\g^{mn}\bar\l =0$, which fixes 11 of the 16 components of
$L^\a$. The remaining 5 components of $L^\a$ will be gauge-fixed to zero using
the invariance of \twistorg\ in which the gauge parameter $f^\a$ is also constrained
to satisfy 
\eqn\fconstraint{f \g^{mn} \bar\l =0.}

Finally, the gauge parameter $g_\a$ of \gaugel\ can be used to gauge-fix the Lagrange
multipler $\bar\L_\a$ to satisfy
\eqn\gaugeLbar{\bar\L_\a = \epsilon \bar\l_\a}
where $\epsilon$ is a constant. Note that $\bar\L_\a$ cannot be gauge-fixed to zero
since it is a pure spinor taking values in ${{SO(10)}\over{U(5)}}\times C^*$. In the gauge
of \gaugeLbar, $g_\a$ satisfies the constraint $g\g^m\bar\l=0$.

One can now follow the standard BRST procedure where the gauge parameters $f^\a$
and $g_\a$ are interpreted as fermionic ghosts, and fermionic
antighosts $m_\a$ and $n^\a$ are introduced due to the gauge-fixing of the Lagrange
multipliers $L^\a$ and $\bar\L_\a$. But because the
Virasoro constraint $T$ can be expressed in terms of $C_\a$ and $\l^\a$ as in \lc, there is no
need to introduce the usual Virasoro ghost and antighost, $c$ and $b$, from gauge-fixing
the reparameterization invariance.\foot{If desired, one can treat the
invariances generated by $T$ as independent symmetries if one also includes
the gauge-for-gauge invariances implied by the relation of \lc.  In this case, the
gauge-fixing procedure will generate the usual fermionic $(b,c)$ Virasoro ghosts of conformal
wieght $(2,-1)$ together with a set of bosonic ghost-for-ghosts $(\beta,\gamma)$
which also carry conformal weight $(2,-1)$. Although it will not be verified here,
it is expected that these ghosts and ghost-for-ghosts will
contribute to the BRST operator the terms
\eqn\nonm{Q = Q_0 + \int dz [\g (b-B) + c(T - b\p c - \b\p\g - \p(\b\g)]}
where $Q_0 = \int dz (\l^a d_\a + \bar w^\a r_\a)$ is the usual non-minimal
pure spinor BRST operator and
\eqn\Bdef{B = d_\a {{(\p x^m +\half\theta\g^m\p\theta) (\g_m \lb)^\a}\over{(\l\lb)}} + 
\p \t^\a {{(\l\g_{mn} w)(\g^{mn}\lb)_\a +2 (\l w)\lb_\a}\over{8(\l\lb)}} + ...}
is the composite ghost satisfying $\{Q_0, B\} = T$ with $...$ denoting terms depending
on the non-minimal variables $(r_\a, s^\a)$. Note that $Q = e^{U}(Q_0 + \g b) e^{-U}$
where $U = \int dz(c B - c \p c \beta)$ and that the structure of $B$ in \Bdef\ resembles
the structure of \lc.}

The resulting gauge-fixed action is 
\eqn\gfa{S = S_0 - \int d^2 z~ Q (m_\a L^\a + n^\a (\Lb_\a - \epsilon \lb_\a) ) - \int d^2 z~
\widehat Q (\widehat m_\ah \widehat L^\ah + \widehat n^\ah (\widehat{\bar\L}_\ah - \widehat\epsilon\widehat{\bar\l}_\ah))}
$$= \int d^2 z [\half \p x^m \bar\p x_m + w_\a \bar\p \l^\a + \widehat w_\ah \p \widehat\l^\ah $$
$$
+ L^\a C_\a + \bar\Lambda_\a \l^\a + \widehat L^\ah \widehat C_\ah + \widehat {\bar\Lambda}_\ah \lh^\ah +{1\over 4} (L\g^m\l)(\widehat L\g_m\lh)  $$
$$- M_\a L^\a - N^\a (\Lb_\a - \epsilon \lb_\a) + m_\a \bar\p f^\a + n^\a ( \bar\p g_\a
+ {1\over 8} (\g^m L)_\a \p(\l\g_m f) -{1\over 8}(\g^m f)\p (\l\g^m L))$$
$$-\widehat M_\ah \widehat L^\ah - \widehat N^\ah (\widehat{\bar\Lambda}_\ah -\widehat \epsilon \widehat{\bar\l}_\ah) + \widehat m_\ah \p \widehat f^\ah + \widehat n^\ah (\p \widehat g_\ah + {1\over 8} (\g^m \widehat L)_\ah \bar\p(\lh\g_m \widehat f) -{1\over 8}(\g^m \widehat f)\bar\p (\lh\g^m \widehat L))]$$
where $S_0$ is the action of \action\ in conformal gauge, $(M_\a, N^\a)$ are bosonic Nakanishi-Lautrup fields associated with the gauge-fixing
of $L^\a$ and $\Lb_\a$, and 
\eqn\brstone{Q = \int dz [\l^\a g_\a + C_\a f^\a -{1\over 8}
(n\g^m f)\p(\l\g_m f) ] }
is the BRST operator which generates the BRST transformations
\eqn\brstt{ Q x^m = \half \l\g^m f, \quad Q w_\a = g_\a + ..., \quad Q L^\a =\bar\p f^\a,\quad
Q f^\a = 0,}
$$ Q g_\a = -{1\over 8} (\g^m f)_\a \p(\l\g_m f),\quad
 \quad Q m_\a = M_\a, \quad Q n^\a = N^\a,$$
$$Q \Lb_\a =\bar\p g_\a +{1\over{16(\l\Lb)}}(\g^m\g^n\Lb)_\a [\p(\l\g_m f)(\l\g_n L) -\p(\l\g_m L)(\l\g_n f)].$$
Since $L\g^{mn}\lb=0$ and $\Lb\g^m\Lb=0$ imply that only
5 components of $L^\a$ and 11 components of $\Lb_\a$ are independent, one can
choose the antighosts and Nakanashi-Lautrup fields to satisfy the constraints
\eqn\antigc{\l\g^{mn} m = \l\g^{mn} M =0 \quad {\rm and} \quad  \l\g^m n = \l\g^m N =0.}

\newsec{ Gauge-Fixed Pure Spinor Formalism}

After integrating out the Lagrange multipliers and Nakanashi-Lautrup fields, one obtains
the equations
\eqn\nla{L^\a = 0, \quad  \Lb_\a - \epsilon\lb_\a =0, }
$$M_\a = C_\a +{1\over 8}
(\g^m n)_\a \p (\l\g_m f) +{1\over 8} (\g^m \l)_\a \p (n\g_m f),\quad N^\a =\l^\a, $$
and the action
\eqn\actiontwo{ S = \int d^2 z (\half \p x^m \bar\p x_m + w_\a \bar\p \l^\a + \widehat w_\ah \p \widehat\l^\ah 
+ m_\a \bar\p f^\a + n^\a \bar\p g_\a 
+ \widehat m_\ah \p \widehat f^\ah +\widehat n^\ah \p \widehat g_\ah).}

Since $\lb_\a$ appears in the action of \actiontwo\ and in the BRST operator of
\brstone\ through the constraints on the ghosts and antighosts, this gauge fixing naively appears to depend on the choice of patch ${\cal O}_\a$.
However, after a cleverly chosen field redefinition, all dependence on $\lb_\a$ can
be eliminated from the action and the BRST operator, and one can take the limit
$\epsilon \to 0$ in the gauge-fixing condition $\Lb_\a = \epsilon \lb_\a$.

The field redefinition involves defining a new unconstrained fermionic variable $\t^\a$
and its conjugate momentum $p_\a$ in terms of the constrained
variables $(f^\a, g_\a, m_\a, n^\a)$ as
\eqn\tpdef{ \t^\a = f^\a + n^\a \quad{\rm and}\quad 
p_\a = e^R ( g_\a + m_\a ) e^{-R}}
where 
\eqn\simr{R = -{1\over {24}}\int dz [ (n \g^{m} \p n)(n \g_{m} f) +3 (n\g^m \p f)(n\g_m f)].} 
Note that
\eqn\lambdap{\l^\a p_\a = e^R (\l^\a g_\a) e^{-R} = \l^\a g_\a +
{1\over 8}(\l\g^m f)(n\g_m\p n) +{1\over 4} (\l\g^m f)(n\g_m \p f),}
and if one had chosen a different
$\d\Lb_\a$ in \deltag\ which also satisfied \implc, the similarity transformation
$R$ of \simr\ would be modified in a manner to leave the BRST operator invariant when expressed in terms of $\t^\a$ and $p_\a$.

It is easy to verify that all 16 components of $\t^\a$ and $p_\a$ in \tpdef\ are unconstrained since
the 5 independent components of $f^\a$ and $m_\a$ are in different directions from the
11 independent components of $g_\a$ and $n^\a$. However, since $(f^\a,g_\a)$ and
$(m_\a, n^\a)$ are ghosts and antighosts which carry conventional ghost number $+1$
and $-1$, $\t^\a$ and $p_\a$ of \tpdef\ do not have well-defined ghost number with respect to
the conventional definition. Nevertheless, one can define a new ghost number where
$(x^m, \t^\a, p_\a)$ carry zero ghost number and $(\l^\a, w_\a)$ carry ghost number
$(+1,-1)$. With respect to this new ghost number, the worldsheet action will carry zero
ghost number and the BRST operator will carry $+1$ ghost number as desired.

After a suitable shift of $w_\a$ to absorb terms proportional to $\bar\p \l^\a$, the action and BRST operator of \actiontwo\ and \brstone\
can be simply expressed in terms of $\t^\a$ and $p_\a$ of \tpdef\ as
\eqn\actionthree{ S = \int d^2 z ( \half\p x^m \bar\p x_m + w_\a \bar\p \l^\a + \widehat w_\ah \p \widehat\l^\ah 
+ p_\a \bar\p \t^\a +\widehat p_\ah \p{\widehat\t}^\ah),}
\eqn\brsttwo{ Q= \int dz (\l^\a p_\a -\half \p x^m (\l\g_m\t) -{1\over 8} (\l\g^m\t)(\t\g_m \p\t)) = \int dz \l^\a d_\a}
where 
\eqn\ddef{ d_\a = p_\a-\half \p x^m (\g_m\t)_\a -{1\over 8} (\g^m\t)_\a (\t\g_m \p\t)}
is the spacetime supersymmetric Green-Schwarz-Siegel constraint.
So one recovers the spacetime supersymmetric
gauge-fixed action and BRST operator of the ``minimal'' pure
spinor formalism which is manifestly independent of the choce of $\lb_\a$.

\newsec{Gauge-Fixed Non-Minimal Pure Spinor Formalism}

To obtain the non-minimal pure spinor formalism \nonmin\ from gauge fixing, one
upgrades the constant pure spinor $\lb_\a$ to a worldsheet variable and constrains
its conjugate momentum $\bar w^\a$ to vanish by adding the term 
\eqn\nmterm{ \int d^2 z (\bar w^\a \bar\nabla \bar\l_\a + \bar w^\a H_\a)}
to the action of \actionthree\ where $H_\a$ is a Lagrange multiplier for the constraint
$\bar w^\a =0$. Since only 11 components of $\bar w^\a$ are independent, the Lagrange multiplier needs to be constrained to
satisfy $H\g^m \lb =0$.

When expressed in terms of $\t^\a$ and $p_\a$, the action of \actionthree\ is independent
of $\lb_\a$ in the limit where the constant $\epsilon$ of $\Lb_\a = \epsilon \lb_\a$ is taken
to zero. To obtain the gauge-fixed nonminimal formalism, one leaves $\epsilon$ nonzero
and defines the non-minimal contribution to the BRST transformations of \brstt\
as 
\eqn\brstnm{Q \lb_\a= - r_\a, \quad Q H_\a = \bar\nabla r_\a,\quad Q s^\a = S^\a,}
where $r_\a$ is the fermionic ghost constrained to satisfy $r\g^m\lb=0$, and
$s^\a$ and $S^\a$ are the antighost and Nakanishi-Lautrup field associated to $H_\a$. Note
that since $\t^\a$ and $p_\a$ are defined to be independent of $\lb_\a$, their BRST transformations do not involve $r_\a$ and
are
\eqn\btp{Q\t^\a = \l^\a, \quad Q p_\a = C_\a -{1\over 8}[ (\t\g^m\p\t)(\g_m \l)_\a -
\p(\l\g^m\t)(\g_m\t)_\a -2 (\l\g^m \t)(\g_m\p\t)_\a ].}

After gauge-fixing $H_\a =0$, the resulting gauge-fixed action and BRST operator are
\eqn\actionfour{ S = \int d^2 z [\half \p x^m \bar\p x_m + w_\a \bar\p \l^\a + \widehat w_\ah \p \widehat\l^\ah 
+ p_\a \bar\p \t^\a +\widehat p_\ah \p{\widehat\t}^\ah}
$$ + \bar w^\a \bar\p \l_\a + \widehat{\bar w}^\ah \p\lh_\ah + s^\a \bar\p r_\a + \widehat s^\ah \p \widehat r_\ah
+ \epsilon (\l^\a\lb_\a + \t^\a r_\a) + \widehat\epsilon (\lh^\ah \widehat{\lb}_\ah + \widehat\t^\ah \widehat r_\ah) ],$$
\eqn\brstnm{ Q = \int dz (\l^\a d_\a + \bar w^\a r_\a),}
where the term $\epsilon (\l^\a\lb_\a + \t^\a r_\a) = \epsilon (\l^\a\lb_\a + n^\a r_\a)$ in
\actionfour\ comes from the
gauge-fixing term
$- Q (n^\a (\Lb_\a - \e \lb_\a))$ in \gfa.
Equations \actionfour\ and \brstnm\ are the gauge-fixed action and BRST operator of the non-minimal pure spinor
formalism \nonmin\ where the term $e^{-\int d^2 z \epsilon (\l^\a \lb_\a + \t^\a r_\a)}$ in
$e^{-S}$ plays the role of a BRST-invariant regulator for integration over the zero modes
of the pure spinors.

\newsec{Generalization to Curved Backgrounds}

The natural conjecture for generalizing the worldsheet reparameterization invariant action of 
\action\ to a curved Type II target-space background is
\eqn\actcurved{S = \int d^2 z (\det e) [\half (g_{mn}(x) + b_{mn}(x)) \nabla x^m \bar\nabla x^n
+ w_\a \bar\nabla \l^\a + \widehat w_\ah \nabla \lh^\ah}
$$ +\Omega_m{}^{np}(x) \bar\nabla x^m  (w\g_{np} \l) +  \widehat\Omega_m{}^{np}(x)\nabla x^m
(\widehat w\g_{np}\widehat\l) + R_{mnpq}(x)(w\g^{mn}\l)(\widehat w\g^{pq}\lh)$$
$$+ L^\a C_\a + \bar\Lambda_\a \l^\a + \widehat L^\ah \widehat C_\ah + \widehat {\bar\Lambda}_\ah \lh^\ah +{1\over 4} (L\g^m\l)(\widehat L\g_m\lh) ] $$
where $g_{mn}(x)$ and $b_{mn}(x)$ are the target-space metric and Kalb-Ramond field,
\eqn\connections{\Omega_m{}^{np} = \Gamma_m{}^{np} + H_m{}^{np} \quad{\rm and}\quad
\widehat\Omega_m{}^{np} = \Gamma_m{}^{np} - H_m{}^{np}} 
are the left and right-moving connections constructed as in the RNS action from the Christoffel connection $\Gamma_m{}^{np}$ and the torsion $H_{mnp} = \p_{[m} B_{np]}$, $R_{mnpq}$ is
the Riemann curvature tensor, $\g^m_{\a\b} = E^m_a(x) \g^a_{\a\b}$ where $a=0$ to 9 is
a tangent-space index and $E^m_a$ is the target-space
vierbein satisfying $\eta^{ab} E^m_a E^n_b = g^{mn}$, and 
\eqn\twistorcurve{C_\a =-\half \nabla x^m (\g_m \l)_\a \quad{\rm and}\quad
\widehat C_\ah =-\half \bar\nabla x^m (\g_m \lh)_\ah } 
are the twistor-like constraints in the curved background. 

Surprisingly, the action of \actcurved\ has the same structure as the RNS worldsheet action if one replaces the left and right-moving
pure spinor Lorentz currents $(w\g^{mn}\l)$ and $(\widehat w\g^{mn}\lh)$ in \actcurved\ with the
left and right-moving RNS Lorentz currents $\psi^m\psi^n$ and $\widehat\psi^m\widehat\psi^n$
and replaces the Lagrange multipliers $(L\g^m \l)$ and $(\widehat L\g^m \lh)$ in \actcurved\
with $\xi \psi^m$ and $\widehat\xi\widehat\psi^m$ where $\xi$ and $\widehat\xi$ are the RNS worldsheet
gravitini and $\psi^m$ and $\widehat\psi^m$ are the RNS fermionic vectors. Just as the
structure of the RNS action is determined by worldsheet supersymmetry, the
structure of \actcurved\ is determined by the requirement that 
$C_\a$ and $\widehat C_\ah$ in \twistorcurve\ generate symmetries of the action.

Although the Ramond-Ramond background fields do not appear in \actcurved, one expects
that consistency of the gauge-fixing procedure will require that they appear in both the
BRST transformations and in the gauge-fixed action. To be more specific, one needs to follow the procedure of \tpdef\ and construct $(\t^\a, p_\a)$ and $(\widehat\t^\ah, \widehat p_\ah)$ variables in terms of the Fadeev-Popov
ghosts and antighosts such that $(\t^\a, p_\a)$ and $(\widehat\t^\ah, \widehat p_\ah)$ are independent
of the choice of patch of pure spinor space.
It is expected that this construction will necessarily involve the Ramond-Ramond background
fields and will imply equations of motion for all of the background fields. So instead of
obtaining the equations of motion for the NS-NS background fields from quantum worldsheet superconformal invariance as in the RNS formalism, it is conjectured that the equations of motion for all of the background supergravity fields (including the Ramond-Ramond fields) will be obtained in this formalism
by requiring
that the gauge-fixed action and BRST operator are independent of the choice of patch of pure spinor space.

For example, for the Ramond-Ramond plane-wave background, the classical action of \actcurved\
is \eqn\ppw{S = \int d^2 z (det e) [\half  \nabla x^m \bar\nabla x_m +\half \mu^2 (\nabla x^+)(\bar\nabla x^+)
x^j x^j 
+ w_\a \bar\nabla \l^\a + \widehat w_\ah \nabla \lh^\ah}
$$ + \mu^2 (x^j \bar\nabla x^+  (w\g^{j+} \l) + x^j \nabla x^+ (\widehat w\g^{j+}\widehat\l) + 
(w\g^{j+}\l)(\widehat w\g^{j+}\lh) )$$
$$
+L^\a C_\a + \bar\Lambda_\a \l^\a + \widehat L^\ah \widehat C_\ah + \widehat {\bar\Lambda}_\ah \lh^\ah +{1\over 4} (L\g^m\l)(\widehat L\g_m\lh) ] $$
where $j=1$ to 8, $x^\pm = x^0 \pm x^9$, and $\mu^2$ is the nonzero component $R_{+j+j}$ of the curvature. Since the constraints $C_\a$ and $\widehat C_\ah$ of \twistorcurve\ are classically conserved, the action of \ppw\ is invariant under local symmetries analogous to the flat background symmetries of  \twistorg\ and \deltag. But combining the fermionic ghosts and antighosts for these symmetries into unconstrained patch-independent variables, $(\t^\a, p_\a)$ and $(\widehat\t^\ah, \widehat p_\ah)$, is expected to be more complicated than in \tpdef\ and to require Ramond-Ramond coupling terms such as $\mu\int d^2 z (p \g^{+1234} \widehat p)$ in the action. The complete consistency of this gauge-fixing procedure is expected to lead to the conformally invariant pure spinor action for the plane-wave background of \ref\planewaveb{N. Berkovits, ``Conformal field theory for the superstring in a Ramond-Ramond plane wave background,''
JHEP {\bf 0204}, 037 (2002).
[hep-th/0203248].}. 

For a general curved background, the gauge-fixing procedure of section 4 and construction of patch-independent $(\t^\a, p_\a)$ and $(\widehat\t^\ah, \widehat p_\ah)$ variables is expected to imply a gauge-fixed action and BRST operator which coincides with the pure spinor worldsheet action and BRST operator of \BH
\eqn\howeact{ S= \int d^2 z [ (G_{MN}(x,\t,\widehat\t) + B_{MN}(x,\t,\widehat\t)) \p Z^M \bar\p Z^N + ...],}
\eqn\howebrst{Q = \int dz ~\l^\a d_\a , \quad \widehat Q = \int d\bar z ~\lh^\ah \widehat d_\ah,}
where $[G_{MN}, B_{MN}, ..]$ are the Type II supergravity superfields described in \BH,
$Z^M = (x^m, \t^\a, \widehat\t^\ah)$ are the N=2 d=10 superspace variables, and $p_\a$
and $\widehat p_\ah$ are the canonical momentum variables for $\t^\a$ and $\widehat\t^\ah$ defined by
\eqn\pdef{p_\a = d_\a - B_{\a M}(\p Z^M - \bar\p Z^M) - \Omega_\a{}^{mn} (\l\g_{mn} w)
- \widehat\Omega_\a{}^{mn}(\lh\g_{mn}\widehat w) , }
$$\widehat p_\ah =  \widehat d_\ah - B_{\ah M}(\p Z^M - \bar\p Z^M) - \Omega_\ah{}^{mn} (\l\g_{mn} w)
- \widehat\Omega_\ah{}^{mn}(\lh\g_{mn}\widehat w). $$

It would of course be very important to verify these conjectures for the curved Type II supergravity background. The first step would be to study the physical states in an open string background which should include both the super-Yang-Mills gluon and gluino. Since the gluino vertex operator is fermionic, it is absent from the reparameterization invariant action which only depends
on bosonic worldsheet variables. This means that
one should find non-trivial BRST cohomology at nonzero ghost number using the conventional
definition of ghost number where the ghosts $(f^\a, g_\a)$ and antighosts 
$(m_\a, n^\a)$ carry ghost number $+1$ and $-1$. After understanding how this works for
the open superstring, it should be straightforward to generalize to the Type II superstring
by taking the left-right product of two open superstrings.

\vskip 1cm
{\bf Acknowledgments:}
I would like to thank Sergei Cherkis, Andrei Mikhailov, Warren Siegel, Cumrun Vafa, Edward Witten, and especially Nikita Nekrasov for useful discussions over the last several years, and
CNPq grant 300256/94-9 and FAPESP grants 2009/50639-2 and 2011/11973-4
for partial financial support.

\listrefs

\end